\documentclass{aa}
\usepackage{graphicx,txfonts,natbib}
\bibpunct{(}{)}{;}{a}{}{,}

\newcommand{\gtap}{\mathrel{\hbox{\rlap{\lower.55ex \hbox {$\sim$}}
                   \kern-.3em \raise.4ex \hbox{$>$}}}}
\newcommand{\ltap}{\mathrel{\hbox{\rlap{\lower.55ex \hbox {$\sim$}}
                   \kern-.3em \raise.4ex \hbox{$<$}}}}

\begin{document}

\title{Creating ultra-compact binaries in globular clusters through stable mass transfer}
\author{M.V. van der Sluys \and F. Verbunt \and O.R. Pols }
\titlerunning{Creating ultra-compact binaries through stable mass transfer}
\offprints {M.V. van der Sluys, \email{sluys@astro.uu.nl}}

\institute{ Astronomical Institute, Princetonplein 5, NL-3584 CC Utrecht,
            the Netherlands,
            {\tt(sluys@astro.uu.nl)},
            {\tt(verbunt@astro.uu.nl)} and
	    {\tt(pols@astro.uu.nl)}
          }

\date{Received / Accepted }

\abstract{

A binary in which a slightly evolved star starts mass transfer to a neutron star can evolve towards
ultra-short orbital periods under the influence of magnetic braking.  This is called magnetic capture.  
We investigate in detail for which initial orbital periods and initial 
donor masses binaries evolve to periods less than 30--40 minutes within the Hubble time.  
We show that only small ranges of initial periods and masses lead to ultra-short periods, and that
for those only a small time interval is spent at ultra-short periods.
Consequently, only a very small fraction of any population of X-ray 
binaries is expected to be observed at ultra-short period at any time.  If 2 to 6 of the 13 bright X-ray
sources in globular clusters have an ultra-short period, as suggested by recent observations, their 
formation cannot be explained by the magnetic capture model.


}

\maketitle


\section{Introduction}
\label{sec:intro}

The globular clusters belonging to our Galaxy house thirteen
bright ($L_\mathrm{X} \gtap 10^{35}$\,erg\,s$^{-1}$ in the 0.5--2.5 keV range)
X-ray sources, neutron stars accreting from a low-mass companion.
A surprisingly large fraction of these has
ultra-short orbital periods of less than about 40 minutes, as first noticed by \citet{1996ApJ...472L..97D}.
Two of the five orbital periods known are
11.4\,min and 20.6\,min (or its alias 13.2\,min) for the
sources in \object{NGC\,6624} and  \object{NGC\,6712}, respectively
\citep{1987ApJ...312L..17S,1996MNRAS.282L..37H}.
The orbital periods of eight systems are not known,
but for four of them indirect evidence points to an ultra-short
period.
This evidence consists of the absolute magnitude of the optical
counterpart \citep{1994A&A...290..133V}, of the energy distribution
of the X-ray spectrum \citep{spo+01}, and of the maximum flux 
reached during X-ray bursts \citep{2003A&A...399..663K}.
Collating this evidence, \nocite{vl04} Verbunt \&\ Lewin (2004, their Table\,1) suggest that 
two more sources probably, and two others possibly have ultra-short orbital
periods (in NGC\,1851 and NGC\,6652, and in NGC\,7078 and Terzan\,5,
respectively).  The 43.6\,m period found by \citet{1996ApJ...472L..97D} is not the 
period of the bright X-ray source in NGC\,6652, but of a fainter source \citep{2001ApJ...562..363H}.

Thus both among the known periods and among the suggested periods,
about half of the bright X-ray sources have ultra-short orbital
periods.
This is in marked contrast to the period distribution of bright
X-ray sources in the galactic disk, where only one period much shorter than
40 minutes has been suggested so far~\citep{aph0406465}.

Ultra-short-period binaries with a neutron star can be formed in a
number of ways. An expanding giant star can engulf the neutron star,
which then spirals in to form a binary with the helium-burning core.
If mass transfer starts immediately after spiral-in, the donor is a
helium-burning star \citep{1986A&A...155...51S}, if mass transfer starts
only after a long time, the donor has evolved into a CO white dwarf or
a CO white dwarf with helium mantle \citep{2002A&A...388..546Y}. The
process requires a giant of higher mass than exists in globular
clusters today; but the waiting time between end of the spiral-in and
onset of the mass transfer allows us to observe the mass transfer
stage today of systems formed long ago. Indeed, it has been argued
that this in fact is the dominant formation process for
ultra-short-period binaries in globular clusters 
\citep{1998MNRAS.301...15D, 2000ApJ...532L..47R}. Alternatively, it has been suggested that
in a cluster, a neutron star can also in a collision with a giant
expell its envelope and form a binary with its core (Verbunt 1987). It
is not obvious that this leads to a binary sufficiently close to start
mass transfer within the Hubble time \citep{1991ApJ...377..559R}. A white
dwarf donor implies an expanding orbit, and thus predicts an
increasing orbital period.

Yet another scenario starts from a binary of a neutron star and a main-sequence
star. The evolution of this binary depends critically on the initial orbital
period. When the period is short, mass transfer is driven by loss of
angular momentum, and the orbital period decreases with the donor mass
until a minimum period is reached near 70\,min \citep{1981ApJ...248L..27P}. 
We will call this a converging system.  At the minimum period, the donor becomes
degenerate, and further mass transfer expands the orbit. 
When the
orbital period is long, mass transfer is driven by expansion of the
donor star, and the orbit expands with the donor radius
until the donor has transferred its full envelope \citep{1983ApJ...270..678W}. 
These are diverging systems.
However, for a narrow range of periods loss of
angular momentum can still shrink the orbit for a slightly evolved
donor. Due to its higher helium content, the donor becomes degenerate
at smaller radius, and correspondingly shorter orbital period~\citep{1985PAZh...11..123T}.
Orbital periods shorter than 11\,min can be reached \citep{2002ApJ...565.1107P}. 
These systems therefore converge, but the process may take more than a
Hubble time.
At 11.4\,min, the period derivative may be negative
or positive, depending on whether the system is still on its way
to the period minimum, or has already rebounded. 
We will refer to this scenario as magnetic capture.

The repeated observation that the 11.4\,min period of the bright X-ray
source in NGC\,6624 is decreasing \citep{1993A&A...279L..21V,2001ApJ...563..934C}
would appear to indicate that the system evolved
according to the magnetic capture scenario. However, it is not impossible that
the negative period derivative is only apparent, the consequence
of an acceleration of the binary in our direction, in the gravitational
potential of the innermost part of the globular cluster.
A more accurate position of the (optical counterpart to the) X-ray
binary and a re-determination of the centre of the cluster shows that
the X-ray source is much closer to the cluster centre than was thought
before, and thus increases the probability that the measured period
is affected by acceleration.  Nonetheless, the measurement of a period
decrease is a strong incentive to investigate the magnetic capture
scenario in more detail.

A possible problem with the magnetic capture scenario is suggested by computations
for binaries in the galactic disk, by \citet{1988A&A...191...57P}. 
None of their calculated evolutions lead to periods of about 11 minutes within the Hubble time.
\citet{2002ApJ...565.1107P} do not address this problem explicitly,
but only show the time elapsed since the onset of mass transfer.

In this paper, we address the question under which circumstances
the very short orbital periods observed in NGC\,6624 and NGC\,6712
are reached within the Hubble time, in the magnetic capture scenario 
described above. The parameters that we vary are the initial mass of the
donor star, the initial orbital period (or more or less equivalently,
the orbital period at which mass transfer starts), and the metallicity
of the donor.
In Sect.~\ref{sec:code} we briefly describe the code that we use, and the
algorithms specific to the evolutionary scenario that we study.
In Sect.~\ref{sec:models} we give the results for two specific cases, to
compare with earlier work and to illustrate the possible evolution paths.  We then describe the
expected outcomes for an initial distribution of donor masses
and initial orbital periods in Sect.~\ref{sec:statistics}. We find that orbital periods
of 11.4 and 20.6\,min are possible, but very unlikely in this
scenario.
The consequences of this conclusion are discussed in Sect.~\ref{sec:discussion}.


\section{Binary evolution code}
\label{sec:code}

\subsection{The stellar evolution code}

We calculate our models using the \textsc{STARS} binary stellar evolution
code, originally developed by \citet{1971MNRAS.151..351E,1972MNRAS.156..361E} and with updated input
physics as described in \citet{1995MNRAS.274..964P}. Opacity tables are taken from
\textsc{OPAL}~\citep{1992ApJ...397..717I}, complemented with low-temperature opacities 
from \citet{1994ApJ...437..879A}.

The equations for stellar
structure and composition are solved implicitly and simultaneously, along
with an adaptive mesh-spacing equation. Convective mixing is modelled by a
diffusion equation for each of the composition variables, and we assume a
mixing length ratio $l/H_p = 2.0$.  Convective overshooting is taken into
account as in \citet{1997MNRAS.285..696S}, with a free parameter
$\delta_\mathrm{ov}=0.12$ calibrated against accurate stellar data from
non-interacting binaries \citep{1997MNRAS.285..696S,1997MNRAS.289..869P}.
The helium core mass is defined as the mass coordinate where the hydrogen 
abundance becomes less than 10\%.

We use a version of the code (see \citet{2002ApJ...575..461E},
hereafter EK02) that allows for non-conservative binary evolution, even
though the evolution of only one component star is calculated in
detail. The companion, in our case a neutron star, is treated as a point
mass. With the adaptive mesh, mass loss by stellar winds or by Roche-lobe
overflow (RLOF) in a binary is simply accounted for in the boundary condition for
the mass. Spin-orbit interaction by tides is treated according to the
equilibrium tide theory \citep{1981A&A....99..126H} with a tidal friction timescale as given
by EK02. This is taken into account by solving additional equations for the
moment of inertia $I(r)$, the uniform stellar rotation frequency
$\Omega_\mathrm{rot}$, the orbital angular momentum $J_\mathrm{orb}$ and
the orbital eccentricity $e$. These equations (of which the latter three
are independent of the interior structure and only depend on time) are also
solved implicitly and simultaneously with the usual set of equations, at
little extra computational cost.
The rotation induces a centrifugal potential that influences the stellar 
structure through a reduction of the effective gravity.  The centrifugal 
potential for each mesh point is averaged over a spherical shell.
Rotationally induced mixing is not taken into account in this code.

Unlike EK02, we do not include their model for dynamo-driven mass
loss and magnetic braking. Rather we apply a magnetic braking law without
accompanying mass loss, as discussed in Sect.~\ref{sec:amloss}. This facilitates direct
comparison to previous binary evolution calculations in which similar
assumptions have been made.  Although we follow tidal interaction in
detail, the effect on the current calculation is limited because the short
orbital periods we consider ensure that the orbit is always circularised
and synchronised with the stellar spin.  However, exchange of angular
momentum between spin and orbit is taken into account.

The initial hydrogen and helium abundances of our model stars are a function of the metallicity $Z$:
$X = 0.76 - 3.0Z$ and $Y = 0.24 + 2.0Z$.  In this research we use the metallicities $Z = 0.0001$ (with
$X=0.7597, Y=0.2402$), $Z = 0.002$ (with $X=0.754, Y=0.244$), $Z = 0.01$ (with $X=0.73, Y=0.26$) and $Z = 0.02$ 
(with $X=0.70, Y=0.28$).

\subsection{Angular momentum losses}
\label{sec:amloss}

If the lower mass star in a binary fills its Roche lobe and starts to transfer mass 
to a more massive companion, the orbit will widen, unless there are enough angular
momentum losses to compensate for this effect.  We assume three sources of angular
momentum loss from the system.

The most important source is magnetic braking. Due to magnetic braking, spin angular
momentum is lost from the secondary and eventually, due to the tidal spin-orbit 
coupling, from the orbit.  We use the formula given by \citet{1983ApJ...275..713R}:
\begin{equation}
  \frac{dJ_\mathrm{MB}}{dt} ~=~ -3.8 \times 10^{-30} \, M_2 \, R^4 \, \omega^3 \, \mathrm{dyn~cm} .
  \label{eq:magnbrak}
\end{equation}
Like \citet{2002ApJ...565.1107P}, we apply full magnetic braking when the mass of
the convective envelope of the donor exceeds 2\% of the total mass of the star, and
if $q_\mathrm{conv} < 0.02$ reduce the strength of the magnetic braking in Eq.~\ref{eq:magnbrak} by a factor of
$\exp(1 - 0.02/q_\mathrm{conv})$, where $q_\mathrm{conv}$ is the mass fraction of the convective
envelope of the star.  The fact that the magnetic braking removes angular momentum from the spin of the
star rather than directly from the orbit is different from \citet{2002ApJ...565.1107P}.  The main
difference is that our study takes into account stellar spin at all, which influences the radius
of the star and thus the moment at which Roche-lobe overflow commences.

For short orbital periods, gravitational radiation is a strong source of angular momentum loss.
We use the standard description
\begin{equation}
  \frac{dJ_\mathrm{GR}}{dt} ~=~ -\frac{32}{5} \, \frac{G^{7/2}}{c^5} \, \frac{M_1^2 \, M_2^2 \, \left(M_1 + M_2\right)^{1/2}}{a^{7/2}}
  \label{eq:gravrad}
\end{equation}
\citep{pet84}.

The third way of angular momentum loss from the system is by non-conservative mass
transfer.  We assume that only a fraction $\beta$ of the transferred mass is accreted by the
neutron star. The remainder is lost from the system, carrying away a fraction 
$\alpha$ of the specific angular momentum of the neutron star
\begin{equation}
  \frac{dJ_\mathrm{ML}}{dt} ~=~ - \alpha\left(1-\beta\right) a_1^2 \, \omega \, \dot{M}_2,
  \label{eq:am_ml}
\end{equation}
where $a_1$ is the orbital radius of the neutron star and $\omega$ is the orbital frequency.

To keep the models simple, we applied no regular stellar wind to our models, so
that all mass loss from the system and angular momentum loss due to this result from
the non-conservative mass transfer described above.


\section{Binary models}
\label{sec:models}

\subsection{Calculated grid}
\label{sec:grid}
Using the binary evolution code described in Sect.~\ref{sec:code}, we calculated
an initial grid of models for $Z = 0.01$, the metallicity of NGC\,6624, and Y=0.26.  We choose 
initial masses between 0.7 and 1.5\,$M_\odot$ with steps of 0.1\,$M_\odot$, and 
initial periods between 0.50 and 2.75 days, with steps of 0.25 days.
Around the bifurcation period between converging and diverging systems, where the shortest
orbital periods occur, we narrow the steps in P to 0.05 days.  

We specify the bifurcation period more precisely as the longest initial period
that leads to an ultra-short period, {\it within a Hubble time}.  
With this definition, the bifurcation period
corresponds to the initial period of the binary that reaches its minimum period
exactly after a Hubble time.  This extra constraint is needed because
there is no sharp transition between converging and diverging systems, especially 
since every diverging system will eventually converge due to gravitational radiation, 
if given the time.  For instance, the system with an initial secondary mass of 1.1\,$M_\odot$
and an initial period of 0.90 days --- that is shown to run out of the right of 
Fig.~\ref{fig:t-p_1.1} at log\,$P\approx-0.4$ --- does converge to a period of slightly 
more than 5 minutes, but only after almost 32 Gyr.  This system is therefore considered to be diverging.
Since the last part of the converging tracks in Figs.~\ref{fig:t-p_1.0} and \ref{fig:t-p_1.1} is very
steep, a system that reaches an ultra-short minimum period shortly after a Hubble time will usually 
have an orbital period at the Hubble time that is on the order of hours.

The total number of calculations for $Z = 0.01$ is 150;
90 for the initial grid, and 60 for the finer grid. We follow \citet{2002ApJ...565.1107P} 
in choosing $\alpha = 1$ and $\beta = 0.5$ in Eq.~\ref{eq:am_ml}. The orbital evolution 
of the systems with initial masses of 1.0 and 1.1\,$M_\odot$ is displayed in 
Figs.~\ref{fig:t-p_1.0} and \ref{fig:t-p_1.1}.

\subsection{Interpretation of the models}
\begin{figure}
\resizebox{\hsize}{!}{\rotatebox{-90}{\includegraphics{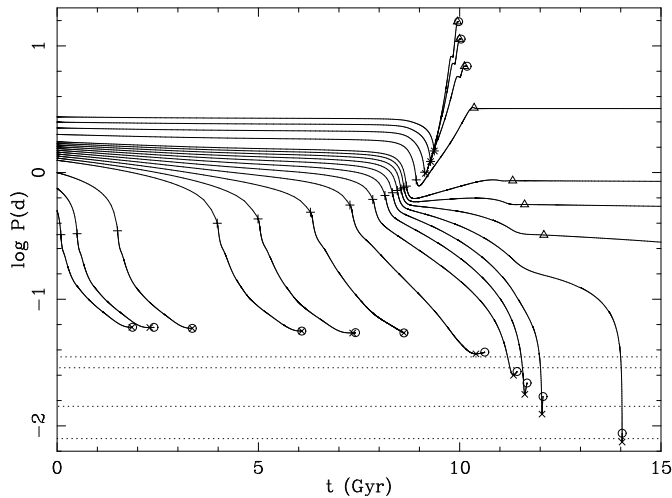}}}
 \caption{
 Evolution of the orbital periods of selected systems with $Z = 0.01$,  
 an initial secondary
 mass of 1.0\,$M_\odot$ and initial periods of 0.50, 0.75, 1.00, 1.25, 1.30, 1.35, 1.40,
 1.45, 1.50, 1.55, 1.60, 1.65, 1.70, 1.75, 2.0, 2.25, 2.5 and 2.75 days.
 The symbols mark special points in the evolution: $+$ marks the start of Roche-lobe 
 overflow (RLOF), $\times$ the minimum period, $\bigtriangleup$ the end of RLOF
 and $\bigcirc$ marks the end of the calculation.  The four dotted horizontal lines show the
 orbital periods of the closest observed LMXBs in globular clusters: 11.4 and 20.6, and
 in the galactic disk: 41 and 50 minutes.
 }
 \label{fig:t-p_1.0}
\end{figure}

Fig.~\ref{fig:t-p_1.0} shows that the models with the shortest initial periods converge
to minimum periods of about 70\,minutes.  After this, the stars become degenerate,
and the orbits expand.  Before the minimum period is reached, the stars become fully
convective, thus mixing all of the star to a homogeneous composition.  These stars
have not yet formed a helium core, but are still a mixture of hydrogen and helium when they
become degenerate.  The stars with larger initial periods have a lower hydrogen abundance
when they reach their minimum period.

For the longest initial periods, the Roche lobe is filled in a later evolution stage and the 
evolutionary time scale is shorter, so that the star expands faster and the mass transfer 
rate is higher.  Because of this, and the fact that the mass ratio is less than 1, 
the angular momentum loss is not strong
enough to shrink the orbit, so that it starts to expand shortly after mass transfer starts.
These stars are sub-giants, and have a compact helium core inside their hydrogen envelopes.
After they have transferred all of this envelope, they shrink and become helium white dwarfs.
The systems with larger initial periods are more evolved when they fill their Roche lobes
and produce more massive white dwarfs. 

In between the smallest and largest initial periods, there are a number of models that reach 
orbital periods that are much shorter than 70\,min.  This happens due to magnetic capture: the
orbital period is reduced strongly under the influence of strong magnetic braking. When magnetic
braking disappears, the orbit is close enough to shrink to ultra-short periods by angular momentum loss due to 
gravitational radiation.  The magnetic captures come from models with a very narrow initial period range.  
The four models with $M_\mathrm{i} = 1.0 M_\odot$ that reach a period less than 40\,minutes, for instance, have initial periods
of 1.45 1.50, 1.55 and 1.60\,days, where the last model reaches the ultra-short period regime
only after 14\,Gyr.  By interpolation, as described later in Sect.~\ref{sec:interpol},
we find that the models that reach a minimum period below 40\,min and within 13.6\,Gyr, have
initial periods in the range 34.5 -- 38.1\,hours.  These stars fill their Roche lobes when their orbital
periods are in the range of 14.3 -- 17.2\,hours. The lowest orbital period reached, by the
system with the initial period of 38.1 hours, is 12.0\,min, after 13.6\,Gyr.

If one draws a vertical line in Fig.~\ref{fig:t-p_1.0} at 11.5\,Gyr (about the age of
the globular clusters), one can imagine that there is a distribution of observable
X-ray binaries at that moment in time.  The lowest orbital period found at that time
is about $10^{-1.75}$\,days, or 25\,minutes.  All models with orbital periods higher than
about 1\,day have stopped mass transfer and will not be visible as X-ray binaries.
Because the lines in Fig.~\ref{fig:t-p_1.0} are steeper at lower periods, it is clear
that the higher periods, around one day, will dominate.

\begin{figure}
\resizebox{\hsize}{!}{\rotatebox{-90}{\includegraphics{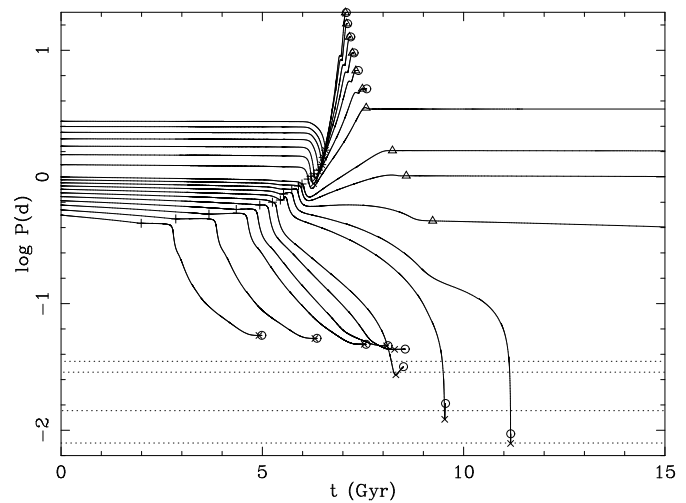}}}
 \caption{
Evolution of the orbital periods of selected systems with $Z = 0.01$,
 an initial secondary
 mass of 1.1\,$M_\odot$ and initial periods of 0.50, 0.55, 0.60, 0.65, 0.70, 0.75, 0.80, 
 0.85, 0.90, 0.95, 1.00, 1.25, 1.50, 1.55, 1.60, 1.65, 1.70, 1.75, 2.0, 2.25, 2.5 and 2.75 days.
 See Fig.~\ref{fig:t-p_1.0} for more details.
 }
 \label{fig:t-p_1.1}
\end{figure}

Figure~\ref{fig:t-p_1.1} shows the same data as Fig.~\ref{fig:t-p_1.0}, but for models
with an initial secondary mass of 1.1\,$M_\odot$.  The results are qualitatively similar,
but the ultra-short period regime is reached from lower initial periods, and after
a shorter period of time.  We find that the models that reach periods lower than 
40\,min before 13.6\,Gyr have initial periods of 18.0 -- 20.9\,hr and fill their Roche lobes
in the period range 15.1 -- 18.2\,hr.  The system with the initial period of 18.0\,hr
reaches 40\,min after 8.3\,Gyr, the system with a 20.9\,hr initial period has the smallest
minimum period: 8.0\,min.

If we again imagine the period distribution at 11.5\,Gyr, but now for Fig.~\ref{fig:t-p_1.1},
we see that the period range that we expect for mass transferring binaries is shifted
downwards in period.  Orbital periods as short as 10.6\,min can now occur, and systems with
periods over 9.5\,hr do not transfer mass anymore at that moment.  With respect to
the tracks in Fig.~\ref{fig:t-p_1.0}, we see that their density is much lower here.  This
is partially due to the fact that we use linear equally spaced periods at a lower initial 
period, so that they are more widely spaced in $\log P$.

\begin{figure}
\resizebox{\hsize}{!}{\rotatebox{-90}{\includegraphics{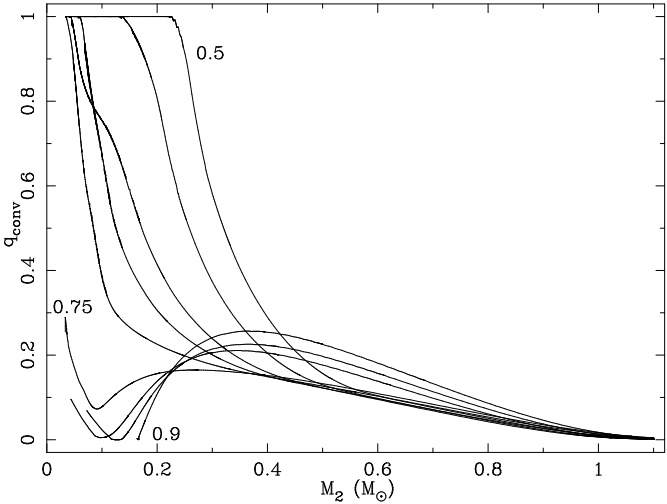}}}
 \caption{
Mass fraction of the convective envelope ($q_\mathrm{conv}$) as a function of the total mass of the donor,
for the models with the shortest 9 initial orbital periods in Fig.~\ref{fig:t-p_1.1}. The numbers in the plot
give the initial periods in days for that line.  As evolution proceeds 
towards lower donor masses, the mass faction of the convective envelope increases.  For the 5 models with
initial periods between 0.5 and 0.7\,d, the total mass at which the star becomes fully convective is 
anti-correlated with the initial period.  At initial periods of 0.75\,d and longer, the initial increase
of the mass fraction of the convective envelope is followed by a decrease. 
 }
 \label{fig:conv}
\end{figure}

Figure~\ref{fig:conv} illustrates the evolution of the convective envelope of a 1.1\,$M_\odot$ star
for the grid models with initial periods between 0.5 and 0.9\,d.  Looking at the models in the order 
of increasing initial period we find that in the first five the stars become fully convective at 
decreasing total masses.
The first model that evolves towards ultra-short periods, with an initial 
period of 0.75\,d is also the first model
in which the donor never becomes fully convective: an initial increase of the mass fraction
of the convective envelope is followed by a decrease.  For initial periods of 0.85\,d and 0.9\,d
the convective envelope disappears completely.  The general trend with increasing initial period
that is visible in Fig.~\ref{fig:conv}, is the consequence of an increasing helium abundance in the core.
The cores with a higher helium abundance tend to be hotter and thus more stable against convection. The
absence of convection in the core in turn keeps the helium abundance high.  
The third model, with an initial period of 0.6\,d shows a track that is slightly different from those of 
the neighbouring models. This model becomes almost fully convective, but the central $10^{-4}\,M_\odot$ does not,
and as a consequence the mixing from the core to the surface is suppressed. We have repeated this calculation 
with a slightly different convective mixing efficiency and find the same results. 

\begin{table*}
\begin{tabular}{llllllllllllll}
  $P_\mathrm{i}$ (d) & $P_\mathrm{rlof}$ (h) & $P_\mathrm{min}$ (m) & $t$ (Gyr) & $\dot{M}_\mathrm{tr}$ & $M_2$ (M$_\odot)$ & $\log L/L_\odot$ & 
  $\log T_\mathrm{eff}$ & $\log T_\mathrm{c}$ & $\log \rho_\mathrm{c}$ & $\log X_\mathrm{c}$ & $\log Y_\mathrm{c}$ & $\log X_\mathrm{s}$ & $\log Y_\mathrm{s}$ \\
 \hline
  0.50 & 10.3  & 80.7   &  4.92 & -10.31 & 0.060 & -3.64 &  3.33 &  6.44 &  2.53 & -0.23     & -0.40 & -0.23 & -0.40  \\
  0.55 & 11.2  & 76.3   &  6.31 & -10.39 & 0.052 & -3.75 &  3.32 &  6.45 &  2.57 & -0.31     & -0.30 & -0.31 & -0.30  \\
  0.60 & 12.2  & 68.7   &  7.53 & -10.49 & 0.042 & -3.85 &  3.32 &  6.54 &  2.81 & -1.46     & -0.02 & -0.53 & -0.16  \\
  0.65 & 13.3  & 66.8   &  8.09 & -10.31 & 0.038 & -3.96 &  3.31 &  6.54 &  2.68 & -0.64     & -0.12 & -0.64 & -0.12  \\
  0.70 & 14.4  & 62.7   &  8.28 & -10.36 & 0.043 & -3.67 &  3.38 &  6.72 &  2.82 & -1.32     & -0.03 & -0.89 & -0.07  \\
 & & & & & & & & & & & & & \\
  0.75 & 15.2  & 39.5   &  8.32 &  -9.67 & 0.056 & -3.46 &  3.48 &  6.93 &  3.43 & -2.34     & -0.01 & -1.24 & -0.03  \\
  0.80 & 15.8  & 17.6   &  9.53 &  -8.53 & 0.074 & -3.97 &  3.45 &  7.01 &  4.08 & -$\infty$ &  0.00 & -1.51 & -0.02  \\
  0.85 & 17.6  & 11.3   & 11.17 &  -7.76 & 0.101 & -4.15 &  3.45 &  7.10 &  4.43 & -$\infty$ &  0.00 & -1.84 & -0.01  \\
  0.90 & 19.1  &  5.1   & 31.85 &  -6.62 & 0.164 & -4.89 &  3.34 &  6.81 &  5.05 & -$\infty$ &  0.00 & -1.10 & -0.04  
\end{tabular}
\caption{Properties for the donor stars of some of our grid models with $Z = 0.01$ and $M_\mathrm{i} = 1.1 M_\odot$ at 
their period minimum.  The first three columns list the orbital period initially (at the ZAMS) in days, at Roche-lobe
overflow ($P_\mathrm{rlof}$) in hours and the minimum period ($P_\mathrm{min}$) in minutes. 
The next 11 columns show stellar properties at $P_\mathrm{min}$: the age of the donor (since ZAMS), the logarithm of the mass transfer rate
(expressed in $M_\odot\, \mathrm{yr}^{-1}$), the mass and luminosity of the donor, the logarithms of the effective temperature,
the core temperature (both in K) and the central density (in g\,cm$^{-3}$), and the last four columns show the logarithms of 
the core and surface mass fractions of hydrogen and helium.
}
\label{tab:pmin}
\end{table*}

Table~\ref{tab:pmin} lists some properties of the same nine models shown in Fig.~\ref{fig:conv} at their period minimum.
The first five models all have minimum periods more than 1\,h and more than 1\%\ hydrogen in the core at their minimum, 
whereas the cores of the last four models consist for more than 99\%\ of helium.  With decreasing minimum period, the
mass transfer rates increase rapidly and the luminosities of the donors decrease.

\begin{table*}
\begin{tabular}{llllllllllll}
 $P_\mathrm{orb}$ (min) & t (Gyr) &  $\log \dot{M}_\mathrm{tr}$ & $\log -\dot{P}_\mathrm{orb}$   & $M_2$ (M$_\odot$) & $\log L/L_\odot$ & $\log T_\mathrm{eff}$ &  
 log H & log He  & log C  & log N  & log O \\
 \hline
  $P_\mathrm{ZAMS}$  &   0.000 &   ---  & ---    & 1.100 &  0.17 &  3.79 & -0.14 & -0.59 & -2.75 & -3.28 & -2.30 \\
 & & & & & & & & & & & \\
  80.0  &   8.023 & -10.11 & -12.56    & 0.097 & -2.44 &  3.59 & -0.46 & -0.19 & -5.12 & -2.50 & -2.36 \\
  60.0  &   8.147 &  -9.97 & -12.48    & 0.086 & -2.66 &  3.59 & -0.57 & -0.14 & -5.07 & -2.42 & -2.44 \\
  50.0  &   8.205 &  -9.81 & -12.49    & 0.079 & -2.82 &  3.58 & -0.70 & -0.10 & -5.01 & -2.35 & -2.53 \\
  45.0  &   8.236 &  -9.72 & -12.54    & 0.074 & -2.95 &  3.57 & -0.82 & -0.08 & -4.96 & -2.31 & -2.62 \\
  39.5  &   8.317 &  -9.67 & -$\infty$ & 0.056 & -3.46 &  3.48 & -1.24 & -0.03 & -4.90 & -2.23 & -2.88 \\
 & & & & & & & & & & & \\
  40.0  &  11.145 &  -9.92 & -11.86    & 0.124 & -1.64 &  3.88 & -0.48 & -0.18 & -5.22 & -2.50 & -2.36 \\
  30.0  &  11.156 &  -9.51 & -11.68    & 0.122 & -1.87 &  3.86 & -0.51 & -0.17 & -5.23 & -2.49 & -2.36 \\
  20.0  &  11.163 &  -9.06 & -11.41    & 0.120 & -2.46 &  3.77 & -0.57 & -0.14 & -5.18 & -2.48 & -2.37 \\
  15.0  &  11.165 &  -8.53 & -11.26    & 0.117 & -3.21 &  3.63 & -0.68 & -0.11 & -5.11 & -2.47 & -2.39 \\
  11.3  &  11.167 &  -7.76 & -$\infty$ & 0.101 & -4.15 &  3.45 & -1.84 & -0.01 & -4.81 & -2.31 & -2.62
\end{tabular}
\caption{Some properties for two of our grid models with $Z = 0.01$ and $M_\mathrm{i} = 1.1 M_\odot$ at selected orbital periods.  
{\it First row:} Initial (ZAMS) parameters.
{\it Rows 2-6}: The model with $P_\mathrm{i} = 0.75$\,d and $P_\mathrm{min} = 39.5$\,min.  
{\it Rows 7-11}: The model with $P_\mathrm{i} = 0.85$\,d and $P_\mathrm{min} = 11.3$\,min.  
$\dot{M}_\mathrm{tr}$ in column\,3 is expressed in $M_\odot\,yr^{-1}$ and $T_\mathrm{eff}$ in column\,7 in Kelvin. 
The last five columns give the logarithm of the surface mass fractions of the elements described. 
}
\label{tab:properties}
\end{table*}

In Table~\ref{tab:properties} we list some observational properties along the evolutionary tracks 
of two of our grid models with $Z = 0.01$ and $M_\mathrm{i}=1.1\,M_\odot$.

Although we find that it is possible to reach orbital periods below 40\,minutes without 
spiral-in, but due to magnetic capture instead, it seems that one has to select an 
initial period carefully in order to actually do so.   We also find that it is possible 
to construct a model that has a minimum period as low as the observed 11.4\,min in a time
span smaller than the Hubble time.  The question arises, however, what the chances are that
such a system is indeed formed in a population of stars.  In order to quantify this, we will
expand our parameter space to the entire grid we calculated and do statistics on these tracks 
in Sect.~\ref{sec:statistics}.

\subsection{Bifurcation models}

\begin{figure}
\resizebox{\hsize}{!}{\rotatebox{-90}{\includegraphics{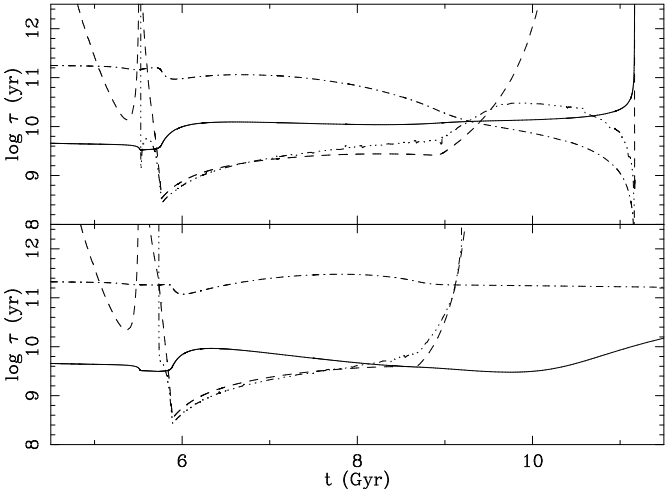}}}
 \caption{
Timescales of the models that bracket the bifurcation period for 1.1\,$M_\odot$. 
 Upper panel (a): model with $P_\mathrm{i}=0.85$\,d.  Lower panel (b): model with $P_\mathrm{i}=0.90$\,d.
 The line styles represent the different timescales: 
 Solid line: nuclear evolution timescale $(M/M_\odot)/(L/L_\odot)\times 10^{10}$\,yr,
 dashes: magnetic braking timescale $J_\mathrm{orb}/\dot{J}_\mathrm{MB}$, 
 dash-dot: gravitational radiation timescale $J_\mathrm{orb}/\dot{J}_\mathrm{GR}$, 
 dash-dot-dot-dot: mass transfer timescale $M/\dot{M}_\mathrm{tr}$.  See the text for a discussion.
 }
 \label{fig:evol_1.1}
\end{figure}

For an initial secondary mass of 1.1\,$M_\odot$, the grid models with initial periods of 0.85\,days and
0.90\,days bracket the bifurcation period.  Some timescales that can explain this difference are shown in
Fig.~\ref{fig:evol_1.1}.
The evolution of both models is rather similar in the beginning, except for the small 
difference in orbital period, that stays about constant during the main sequence.  
The wider system has a larger Roche lobe and thus the donor fills its Roche lobe at a slightly
later stage of its evolution.  At this point, the evolutionary timescale of the donor is shorter
than that in the closer system, and it can form a well defined helium core.
When the envelope outside this core has been reduced by mass transfer to $\simeq 0.03\,M_\odot$, it 
collapses onto the core, mass transfer stops, and magnetic braking disappears before the magnetic 
capture is complete. Gravitational radiation is then
the only term of angular momentum loss and it is not strong enough to shrink the orbit to the ultra-short
period regime within the Hubble time. 

In the closer system, the evolutionary timescale of the donor is slightly larger and its helium core mass
is slightly smaller.  At approximately 9\,Gyr mass transfer has stripped the donor to such extend that hotter 
layers emerge at the surface, the convective envelope of the star becomes very thin and magnetic braking 
is strongly reduced (see the discussion with Fig.~\ref{fig:conv}). Fig.~\ref{fig:evol_1.1} shows that this happens at the moment where the gravitational 
radiation timescale becomes shorter than the evolutionary timescale of the donor, so that angular momentum 
loss remains sufficient to shrink the orbit from the hour to the minute regime.


\section{Statistics}
\label{sec:statistics}

\subsection{Interpolation between models}
\label{sec:interpol}
In order to do statistics on our models, we have to interpolate between the calculated
models to get a time-period track, that gives the orbital period of a system as a function of time,
for a given initial orbital period $P_\mathrm{i}$.

Before we can interpolate between two calculated tracks, we must first divide the 
tracks into similar parts of evolution.  
We choose three parts: i) the part between ZAMS and the beginning of
RLOF, ii) the part between the beginning of RLOF and the moment where the minimum
period ($P_\mathrm{min}$) was reached, and iii) the part between $P_\mathrm{min}$ and the end of
the calculation.  Each of these parts is redistributed into a fixed number of data 
points, equally spaced in the path length of that part and determined by a polynomial 
interpolation of the third degree.  The path length is the integrated track in the t--$\log P$ plane,
and defined as
\begin{equation}
\ell ~=~ \sum_i ~ \sqrt{ \left( \frac{t(i) - t(i-1)}{\Delta t}\right)^2 ~+~ \left(\frac{\log P(i) - \log P(i-1)  }{\Delta \log P}\right)^2  },
\end{equation}
where $\Delta t = t_\mathrm{max} - t_\mathrm{min}$ and $\Delta \log P = \log P_\mathrm{max} - \log P_\mathrm{min}$.
Thus, each part of all tracks contains the same number of points, and each point on these parts marks about the same 
moment in evolution in two different tracks.  

Next, we interpolate between two tracks, to calculate the track for the
given initial period.  Because the tracks differ considerably 
between the shortest and longest initial period, we use linear interpolation between
two adjacent tracks, that are always rather similar.  Each track is thus interpolated
point-by-point between each pair of corresponding points from the two adjacent tracks,
to get the time and the orbital period.  

Once the interpolated track is known, we interpolate within the track, to obtain 
the orbital period at a given moment in time.  For this, we use a polynomial
interpolation of the fourth degree.  For some models the second part of a track consists 
of one point, because the beginning of RLOF marks the minimum period.  For interpolations 
involving this point, we use a third degree polynomial interpolation.

A handful of models crash after they have stopped mass transfer, for instance the models
with the highest initial period in Figs.~\ref{fig:t-p_1.0} and \ref{fig:t-p_1.1}. 
These systems will not give observable X-ray sources, but some of these tracks may be needed for the 
interpolation.  We continued the orbital evolution of the most important of these models 
analytically, under the influence of gravitational radiation only, until the orbit becomes 
so small that the star's Roche lobe touches its surface.  We consider the orbital period at which
mass transfer recommences as the minimum period.
We assume a constant radius of the star since the last converged model, 
which probably means that we overestimate the minimum period a bit in these cases.

\subsection{Results for Z=0.01}
\label{sec:results_z001}

In Sect.~\ref{sec:models}, we have found that we can create LMXBs with periods down to 11 minutes or
perhaps even less, within a Hubble time.  We also saw, however, that one has to select the initial
period carefully to create a model that reaches such a low period, and that the system spends very
little time on this minimum period. In order to investigate how probable it is to {\em observe} 
ultra-compact binaries, we select random points on random tracks like the ones in 
Figs.~\ref{fig:t-p_1.0} and \ref{fig:t-p_1.1} and convert the result into a histogram.  We 
perform this operation in the following way.

\begin{figure}
\resizebox{\hsize}{!}{\rotatebox{-90}{\includegraphics{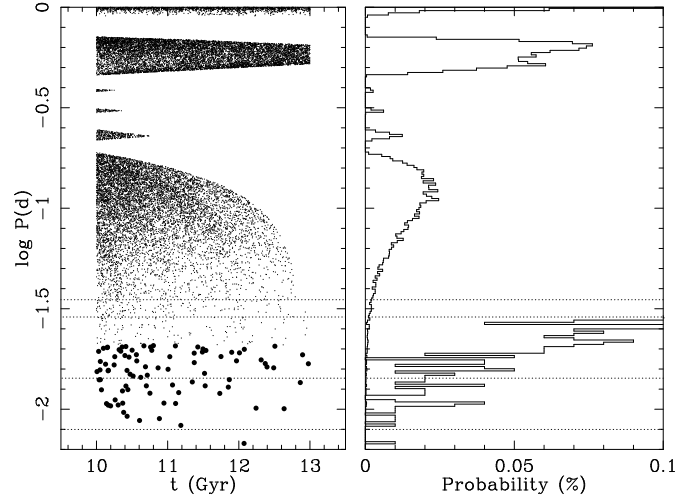}}}
 \caption{ 
 Statistics results for the 1.1\,$M_\odot$ models.
 Left panel (a): Results from the draw of one million random initial periods
 and times.  Each dot represents the orbital period of the selected system
 at the selected time.  Only models that were converging and transferring mass
 at that time were accepted, about 10.5\% of the total number.  The peaks at the
 higher orbital periods are artefacts, caused by interpolation between models
 with and without mass transfer.  Dots below $P = 30\,\mathrm{m}$ 
 are plotted larger for clarity.
 Right panel (b):  A histogram displaying the fraction of systems found at a 
 certain orbital period, at any time between 10 and 13 Gyr. The $\log P$-axis
 was chosen to be vertical, to correspond to the vertical axis in the left
 panel.  The thick line displays the data corresponding to the horizontal
 axis, the thin line is the short-period tail of the same data, 
 multiplied by a factor of 100 in the
 horizontal (probability) direction.  The dotted horizontal lines are the orbital
 periods of the four observed LMXBs mentioned in Fig~\ref{fig:t-p_1.0}.}
 \label{fig:phist_1.1}
\end{figure}

For a fixed initial secondary mass, we draw a random initial period, between 0.50 and 2.75 days, 
from a flat distribution in $\log P$.  We then interpolate the time-period track that 
corresponds to this initial period, using the method described in Sect.~\ref{sec:interpol}.
For each point on this track, an estimate for the mass transfer rate is obtained by interpolating
in the logarithm of the calculated mass transfer rates.  For points without mass transfer, we adapt
a value of $\dot{M}_\mathrm{tr} = 10^{-35}\, M_\odot\, yr^{-1}$, so that we can take its logarithm.
This introduces some irregularities, like the peaks around $\log P\,\mathrm{(d)} = -0.5$ in Fig.~\ref{fig:phist_1.1}, where interpolation
between models with and without mass transfer, and interpolation between converging and diverging
models play a role.  This is usually only the case at orbital periods of several hours or more, and
hence it is of no consequence for the ultra-compact binaries.

Once the time-period track is calculated, we draw a random moment in time, from a linear distribution
between 10 and 13 Gyr, the approximate ages of globular clusters, and interpolate within the 
track to obtain the orbital period at that random moment.  We accept only systems that have not evolved 
beyond their minimum period, firstly because of the negative period derivative measured for the 11.4\,min 
system in NGC\,6624, and secondly because the evolution code we use can generally not calculate far beyond the
period minimum. We also estimate the mass transfer 
rate at that moment, again by interpolating in $\log \dot{M}_\mathrm{tr}$. We reject all systems with a 
mass transfer rate $\dot{M}_\mathrm{tr} < 10^{-20}\, M_\odot\, yr^{-1}$, because it is unlikely that they
have any mass transfer at that moment and will therefore not be an X-ray source.  

If we repeat this procedure many times, we can create a histogram that displays 
the expected distribution of orbital periods of a population of converging LMXBs (with all the same initial 
secondary masses) after 10 - 13\,Gyr.  The results for $1.1\, M_\odot$ and 
$Z = 0.01$ are shown in Fig.~\ref{fig:phist_1.1}.

\begin{figure}
\resizebox{\hsize}{!}{\rotatebox{-90}{\includegraphics{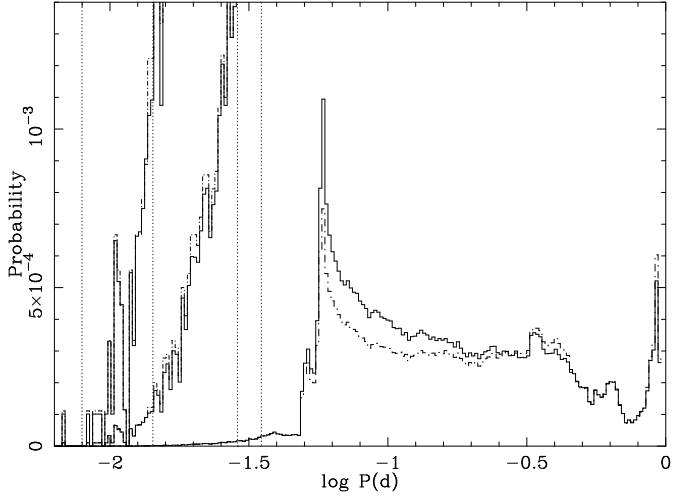}}}
 \caption{ 
 Probability distribution for all models with $Z = 0.01$.  
 The solid line represents the sum of the distributions of the different masses weighed with the Salpeter 
 birth function, the dash-dotted line assumes a flat
 distribution in mass.  The thin lines below $\log P\,\mathrm(d) = -1.3$ and below 
 $\log P\,\mathrm(d) = -1.7$ are the same data, multiplied with 100 and 1000 respectively.
 The four vertical, dotted lines show the orbital periods of the four
 observed LMXBs mentioned in Fig~\ref{fig:t-p_1.0}.
 }
 \label{fig:phist_sum_z01}
\end{figure}

To simulate a population consisting of stars of different masses, one should interpolate 
between the tracks as we did for the period.  The tracks are too different from each other
to ensure correct results.  It would require a large number of extra models to be able to
interpolate between the masses correctly. Instead, we choose to add the period distributions
of the different masses to simulate such a population.  We do this for two different 
assumptions for the mass distribution: the Salpeter birth function, and a flat distribution.
The results are shown in Fig.~\ref{fig:phist_sum_z01}.

We see that there is little difference between the two weighing methods.  This 
assures that although we do not know the initial distribution of the mass, it is of little
influence on this result.  Especially the short-period tails of the distributions 
are almost equal.  In a sample of $10^7$ systems we find one converging system with a period of 
about 11 minutes and 15 systems with a period of 20 minutes.

\subsection{Results for other metallicities}
\label{sec:results_otherz}

\begin{figure}
\resizebox{\hsize}{!}{\rotatebox{-90}{\includegraphics{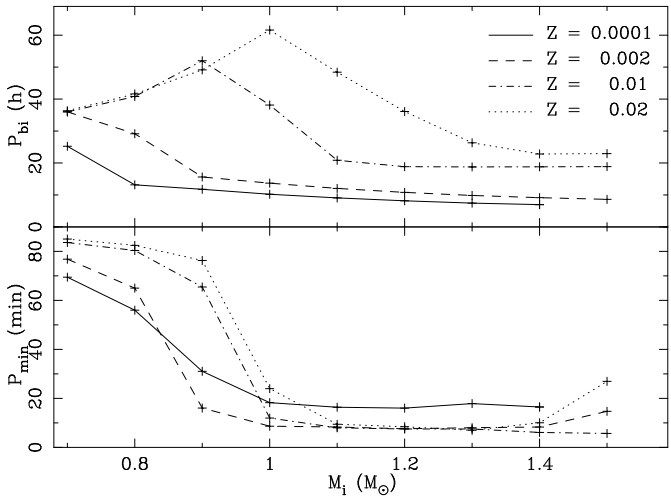}}}
 \caption{ 
 Bifurcation periods and minimum periods
 as a function of the initial mass for the four metallicities.
 {\it Upper panel} (a): The bifurcation period (in hours) between systems that converge and
 systems that do not converge within a Hubble time. 
 {\it Lower panel} (b): The minimum period (in minutes) that can be reached within a Hubble time 
 as a function of the initial secondary mass. The different line styles display the 
 different metallicities, as indicated in the upper panel.  The data point for $Z = 0.0001$, 
 $M_\mathrm{i} = 1.5\,M_\odot$ is missing in both panels, because the bifurcation period for these
 systems is lower than the period at which such a donor fills its Roche lobe at ZAMS.
 }
 \label{fig:bifurc_allz}
\end{figure}

The whole exercise we described in section Sect.~\ref{sec:grid}, \ref{sec:interpol} 
and \ref{sec:results_z001} is also
applied to models for $Z = 0.0001$, $Z = 0.002$, and $Z = 0.02$, in order to see the effect
of metallicity on the expected distributions.  For $Z = 0.02$ we calculate
the same initial grid as we did for $Z = 0.01$, between $P_\mathrm{i} = 0.5 - 2.75$\,days
for $M_\mathrm{i} = 0.7 - 1.3 M_\odot$, but $P_\mathrm{i} = 0.55 - 3.025$\,days for 
$M_\mathrm{i} = 1.4$ and $1.5 M_\odot$, since these stars even at the ZAMS do not fit
in an orbit with $P = 0.5$\,days.  For $Z = 0.002$ we use the same initial mass
range, but it turns out that for $M_\mathrm{i} = 1.0 - 1.5 M_\odot$ the bifurcation
period lies very close to or lower than 0.5 days (see Fig.~\ref{fig:bifurc_allz}).  We therefore shift the minimum
initial period to 0.35\,days for $M_\mathrm{i} = 1.0, 1.4$ and $1.5 M_\odot$, and
to 0.4\,days for $M_\mathrm{i} = 1.1 - 1.3 M_\odot$.  For $Z = 0.0001$, the minimum
initial period is shifted to 0.3\,d for $0.7 - 1.2\,M_\odot$ and even to $0.28\,d$
for $1.3 - 1.5\,M_\odot$.  For $Z = 0.0001$ and $M_\mathrm{i} = 1.5\,M_\odot$, the initial
period at which a ZAMS star fills its Roche lobe is higher than the bifurcation period.
Stars with higher $Z$ have larger radii and often do not fit in these tight orbits.  We shift the upper 
limit for the period range from which we took random values accordingly, so that
the size of the range (in $\log P$) did not change.  Since the bifurcation period for the lower metallicity
models lies lower, we also have to pinpoint better to calculate the interesting models 
around it.  We therefore narrow the grid to steps of 0.01\,d around the
bifurcation period for $Z = 0.002$ and $Z = 0.0001$, and even down to 0.001\,d for the last metallicity.

The bifurcation periods for the different masses are plotted in
Fig.~\ref{fig:bifurc_allz}a.  There is a trend in metallicity in the
sense that the dotted line of $Z = 0.02$ could be moved down and left to fall over that
of $Z = 0.01$ and further to reach that of $Z = 0.002$ and $Z = 0.0001$.  
Fig.~\ref{fig:bifurc_allz}b shows the minimum periods for the systems that have the bifurcation
period for that mass as their initial period.  The trend that is shown can be explained the fact that
low mass stars with a lower metallicity reach the TAMS before the Hubble time and are therefore
eligible for magnetic capture, whereas low mass stars of higher $Z$ do not.

\begin{figure}
\resizebox{\hsize}{!}{\rotatebox{-90}{\includegraphics{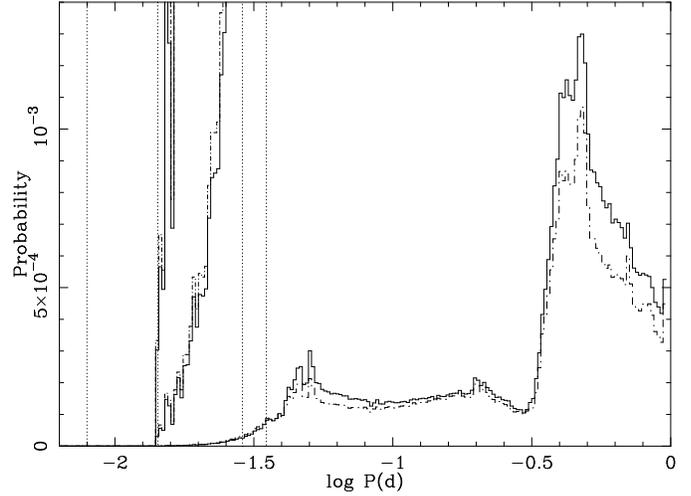}}}
 \caption{ 
 Probability distribution of the orbital
 periods for all models with $Z = 0.0001$.  The characteristics of this plot are the same as in 
 Fig.~\ref{fig:phist_sum_z01}.
 }
 \label{fig:phist_sum_z0001}
\end{figure}

\begin{figure}
\resizebox{\hsize}{!}{\rotatebox{-90}{\includegraphics{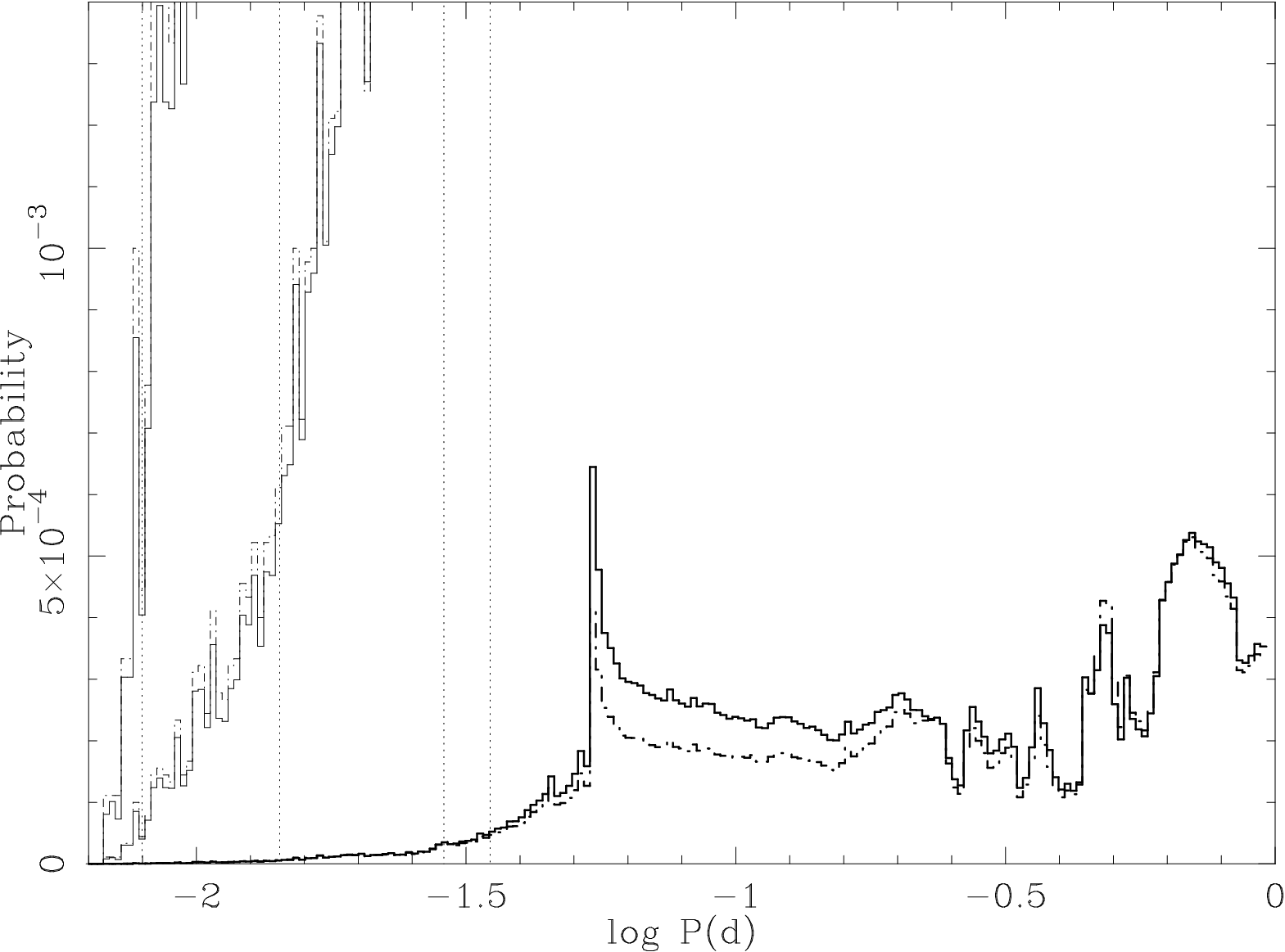}}}
 \caption{ 
 Probability distribution of the orbital
 periods for all models with $Z = 0.002$.  The characteristics of this plot are the same as in 
 Fig.~\ref{fig:phist_sum_z01}.
 }
 \label{fig:phist_sum_z002}
\end{figure}

\begin{figure}
\resizebox{\hsize}{!}{\rotatebox{-90}{\includegraphics{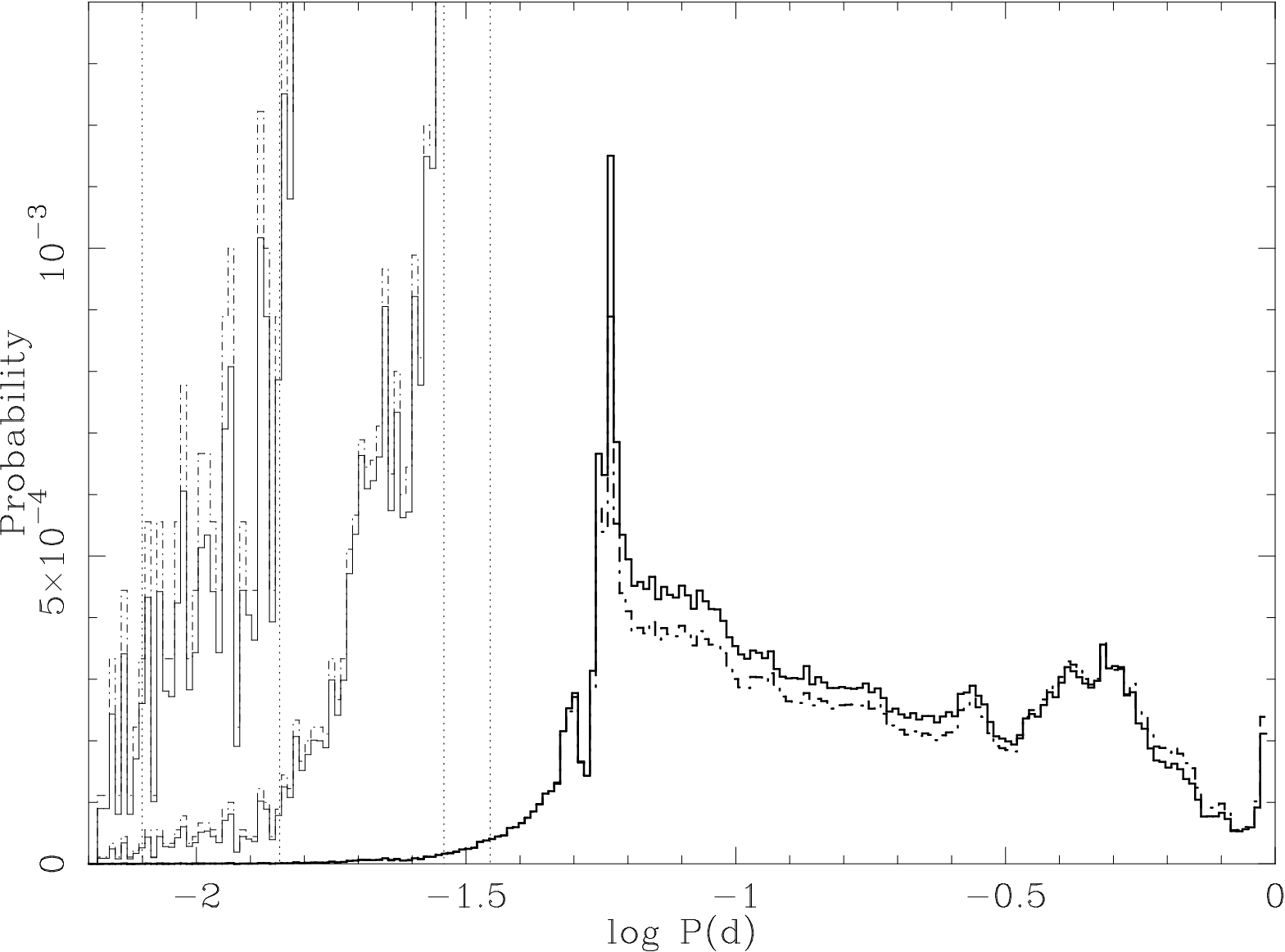}}}
 \caption{ 
 Probability distribution of the orbital
 periods for all models with $Z = 0.02$.  The characteristics of this plot are the same as in 
 Fig.~\ref{fig:phist_sum_z01}.
 }
 \label{fig:phist_sum_z02}
\end{figure}

\begin{figure}
\resizebox{\hsize}{!}{\rotatebox{-90}{\includegraphics{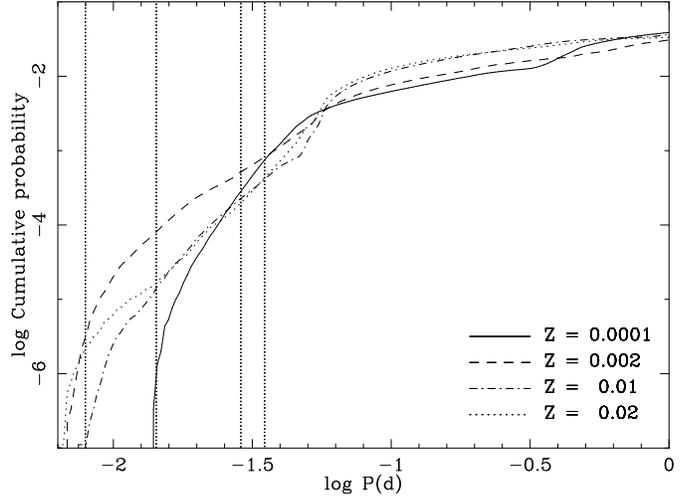}}}
 \caption{ 
Cumulative plot for the distribution of the
 orbital periods for all models and all four metallicities.  The different line styles 
 represent the different metallicities as indicated in the lower right of the plot.
   The height of the lines shows the logarithm of the fraction of all probed systems
 that have an orbital period equal to or lower than the period on the horizontal axis.
 For all lines, a flat initial mass distribution is used.
 The dotted vertical lines show the observed orbital periods mentioned in Fig~\ref{fig:t-p_1.0}.
 }
 \label{fig:pcumul_allz}
\end{figure}

The results of the statistics for $Z = 0.0001$, $Z = 0.002$ and $Z = 0.02$ are plotted in 
Figs.~\ref{fig:phist_sum_z0001}, \ref{fig:phist_sum_z002} and \ref{fig:phist_sum_z02} in the same way
as the results for $Z = 0.01$ in Fig.~\ref{fig:phist_sum_z01}, so that they
can easily be compared.  All four distributions are also plotted in a cumulative plot
in Fig.~\ref{fig:pcumul_allz}, showing the fraction of systems with an orbital period
below some value, so that they can be compared directly.

The most remarkable feature in the three distributions with the higher metallicities is the sharp drop
of the number of predicted systems below $\log P\,\mathrm{(d)} = -1.25$, or about 80\,minutes.
This is due to the systems with low initial mass ($0.7 - 0.9 M_\odot$), that reach their minimum periods 
there because they evolve too slow to reach ultra-short periods before the Hubble time, 
and remain relatively long at this period.  Models with $Z = 0.0001$ evolve more quickly, and although
most models do not reach ultra-short periods, they are substantially lower than 80\,min and can even
reach 31\,min in the case of $M_\mathrm{i} = 0.9\,M_\odot$.  The drop is therefore less sharp for the
lowest metallicity we used.

The lower mass stars dominate in roughly the $\log P$-range 
$-1.25$ -- $-0.6$, as can be seen from the fact that here the solid line for
a Salpeter weighted addition of the masses that favours low mass stars is higher 
than the dash-dotted line for a flat mass distribution.
For the ultra-short periods, there is very little difference between the two weighing
methods, and we can again conclude that the exact initial mass distribution is not important for our
results.

We also see that the lowest possible orbital period for an X-ray binary with $Z = 0.0001$ within the Hubble
time is about a factor two smaller than for the other metallicities.  This is partly due to the fact that
ultra-compact binaries are less likely to be formed for this low metallicity because the initial period
must be chosen more precisely.  However, we find no minimum periods less than 16.0\,min for this 
metallicity.  This has probably to do with the fact that these stars are hotter and thus have a weaker
magnetic field.

In a sample of $10^7$ binaries with $Z = 0.0001$, we expect no converging systems with mass 
transfer and an orbital period of 11.4\,min, and around 5 with a 20.6\,min period 
(Fig.~\ref{fig:phist_sum_z0001}).  For $Z = 0.002$ and $Z = 0.02$, these numbers are 7 systems with an 
11.4\,min period and 60 with a 20.6\,min period and 4 systems with an 11.4\,min period and 10 with a 20.6\,min 
period respectively.

Fig.~\ref{fig:pcumul_allz} shows clearly that there is some difference between the period distributions 
for the different metallicities, the largest difference being the higher period cut-off for the lowest 
orbital periods for $Z = 0.0001$.  The largest differences for the three higher metallicities
are found around 11\,min, (a bit more than an order of magnitude between $Z = 0.01$ and the other
two metallicities) and around 20\,min (less than an order of magnitude between $Z = 0.002$ and the 
others).  Note that the line for $Z = 0.01$ predicts for each system with an orbital period of 11\,min 
about 100 systems with $P_\mathrm{orb} \ltap 20\,\mathrm{m}$.


\section{Discussion}
\label{sec:discussion}

\subsection{The importance of converging evolution for the
formation of ultra-compact binaries}

To understand why the fraction of ultra-compact binaries with decreasing orbital period
in our computations is so small, we note that there are three
main factors contributing to this.
First, only a limited range of initial orbital periods
leads to strongly converging orbital evolution within the
Hubble time, as listed in Table\,\ref{tab:periods}. This range of periods 
varies strongly with donor mass: for $Z = 0.01$ and
for 1.0 and 1.1\,$M_\odot$
the width is about 0.1\,d; but for 1.2 and 1.3\,$M_\odot$ it
is only 0.003\,d. This corresponds to $\sim$5\%\ and $\sim$0.2\%,
respectively, of the range that we consider.
The reason for this rapid decrease is
that the nuclear evolution time scale of the star increases 
much more rapidly with mass than the time scale of magnetic braking.
Thus, at higher stellar mass magnetic capture can only occur for smaller
initial orbital periods.
Second, for each initial orbital period within the range of
converging systems, only a very short time is spent at ultra-short periods
while converging. Thus, the 1.1\,$M_\odot$ system with initial
period of 0.85\,d reaches the 20\,min period after 11.163\,Gyr and
the 11\,min period after 11.167\,Gyr. If we allow a range of
ages of 3\,Gyr, then only 0.1\%\ of these systems will have
an orbital period less than 20\,min {\it and} a negative period derivative.
If we allow also positive period derivatives, the fraction of ultra-compact 
binaries is somewhat higher: as can be seen in Fig.~\ref{fig:t-p_1.1} the evolution 
towards longer period is comparably rapid as the evolution towards shorter period 
close to the minimum period.
Third, as already mentioned, the range of initial periods leading to
converging systems is very small for donors with $M\ge1.2 M_\odot$;
hence only donors in a narrow range of initial masses contribute to
ultra-short period systems.
The combination of these three factors explains why so few ultra-short period 
systems are produced, as already surmised by \citet{1987SvAL...13..328T}.

\begin{table}
\begin{tabular}{ll@{\hspace{0.3cm}}ll@{\hspace{0.3cm}}ll@{\hspace{0.3cm}}ll@{\hspace{0.3cm}}l@{\hspace{-0.cm}}}
 M$_\mathrm{i}$  &    P$_\mathrm{i,1}$  &    P$_\mathrm{i,2}$  &  P$_\mathrm{rlof,1}$  &   P$_\mathrm{rlof,2}$  &  
                     R$_\mathrm{zams}$  &  R$_\mathrm{tams}$  &  P$_\mathrm{zams}$  &  P$_\mathrm{tams}$  \\
 \hline
 1.0  &  1.477  &  1.589  &  0.638  &  0.715  &  0.92  &  1.73  &  0.19  &  2.50 \\
 1.1  &  0.767  &  0.856  &  0.640  &  0.740  &  1.05  &  1.51  &  0.22  &  2.00 \\
 1.2  &  0.753  &  0.756  &  0.686  &  0.689  &  1.18  &  1.71  &  0.26  &  2.37 \\
 1.3  &  0.753  &  0.756  &  0.704  &  0.707  &  1.27  &  2.00  &  0.29  &  2.94 \\
 1.4  &  0.753  &  0.758  &  0.714  &  0.719  &  1.31  &  2.37  &  0.29  &  3.71 \\
 1.5  &  0.752  &  0.763  &  0.717  &  0.728  &  1.33  &  2.65  &  0.29  &  4.32
\end{tabular}
\caption{Comparison between the orbital periods that lead to periods less than 30\,min
within a Hubble time and orbital periods that result from tidal capture with a 1.4\,$M_\odot$ neutron star,
for different secondary masses and $Z = 0.01$. 
Column 1: initial secondary mass, columns 2-3: initial period range that leads to ultra-short periods,
columns 4-5: RLOF-period period range that leads to ultra-short periods, columns 6-7: ZAMS and TAMS radii,
columns 8-9: orbital periods for a circularised binary with capture distances of $1 \times R_\mathrm{zams}$ and $3 \times R_\mathrm{tams}$.
Masses are in $M_\odot$, radii in $R_\odot$ and periods in days.}
\label{tab:periods}
\end{table}

In our computations above we have assumed an initial period
distribution in the range $0.5\,\mathrm{d}\ltap P_b\ltap3\,\mathrm{d}$.  In
the galactic disk, the actual period range extends to much longer
periods, and accordingly our estimates of the fraction of X-ray
binaries that is observed at ultra-short periods are upper bounds, for
systems evolved along the scenario that we compute. This is in
agreement with the absence of large numbers of X-ray binaries with periods much 
less than 40\,minutes, in the galactic disk.  If fact only one such system has recently been discovered;
it may well have formed through a different mechanism, e.g.\ via a double spiral-in at the end of which a
white dwarf becomes the donor of a neutron star \citep{1986A&A...155...51S}.

In globular clusters the binary period distribution is expected to
be different from that in the galactic disk: the widest primordial binaries
are dissolved and close binaries are produced in close stellar encounters.
If the neutron star is exchanged into a primordial binary in a
neutron-star/binary encounter, the period after the encounter
scales with the pre-encounter binary period; in general the
orbit after exchange will be similar in size~\citep{1993ApJ...415..631S}. However, the
range of periods is still expected to be wider than the range that
we have considered in our computations, which therefore give an
upper bound to the fraction of ultra-compact binaries.
If the neutron star is captured tidally, the orbital period
after capture tends to be short. The exact description of tidal
capture is highly uncertain, and we will discuss the simplest
description to provide a reference frame. In this description,
the neutron star captures a main-sequence star if its closest
approach $d$ is within three times the radius $R$ of that star,
i.e.\ $d\le3R$~\citep{1975MNRAS.172P..15F}. The capture rate is linear in $d$; thus one third of the
captures is a direct hit, which completely destroys the main-sequence
star. Capture may lead to a binary if $R\ltap d\le3R$.
The lower bound may in fact be higher, since too close a capture
still does serious damage to the star~\citep{1987A&A...184..164R}. The orbit immediately after
capture is highly eccentric, and after it circularises its semi-major
axis is twice the capture distance: $a_\mathrm{c} \simeq 2d$. Hence orbits formed
by tidal capture have a semi-major axis (after circularisation)
$2R \ltap a_\mathrm{c} \leq 6R$, or with Kepler's law:
\begin{equation}
0.23\,\mathrm{d} \left(R\over R_\odot\right)^{3/2}
 \left(M_\odot\over M+m\right)^{1/2} \ltap P_b \leq
1.20\,\mathrm{d} \left(R\over R_\odot\right)^{3/2}
 \left(M_\odot\over M+m\right)^{1/2}
\label{e:period}\end{equation}

Immediately after the capture, the main-sequence star is highly perturbed,
but after a thermal timescale it may settle on its equilibrium radius, and
continue its evolution.
The range of orbital periods depends on the radius that the star
has when it is captured.
 In general, the period range is bounded below by the
period found by entering twice the zero-age main-sequence radius
into Eq.\,\ref{e:period} and above by entering six times the
terminal-age main-sequence radius (because a star evolved beyond this point
does not evolve towards shorter periods).
In Table\,\ref{tab:periods} we list the period ranges expected in this
simplest description of capture.
Unless the central density of the globular cluster evolves dramatically, the
probability of capture is approximately flat in time.
The period after capture close to the zero-age main sequence
should be compared to the initial binary period in our computations;
the period after capture close to terminal-age main sequence
should be compared to the period of a system close to filling
its Roche lobe.
In either case, we see that capture leads to a period distribution
which covers an appreciable fraction of the period distribution that
we cover in our computations. This means that our conclusion that
only an exceedingly small fraction of all binaries with a
neutron star evolve towards periods less than 30\,min
holds also for tidally captured binaries.

We have taken the simplest description of tidal capture. From the
above argument it is clear that changing the assumptions made
about tidal capture is unlikely to change our conclusion, that
evolution from magnetically driven converging evolution does not
contribute significantly to the population of ultra-compact
binaries. 
Even if tidal capture would miraculously focus the resulting
orbits into the narrow range required for converging evolution,
the fact would remain that each systems spends only a small
fraction of its time converging from 20\,min to 11\,min.

If the binary in NGC\,6624 were the only ultra-short-period binary in a
globular cluster, one could accept an evolutionary scenario with low
probability.  It is thus worthy of note that our statistical argument
depends critically on the observation that the 20.6\,min (or 13.2\,min)
period of the binary in NGC\,6712 is real. So far, this period has
been measured only once in a single HST data set, and an independent
new measurement is very desirable, to exclude
definitely that the first measurement of a significant
periodicity is a statistical fluke.

\subsection{Comparison to Pylyser \& Savonije}
\begin{figure}
\resizebox{\hsize}{!}{\rotatebox{-90}{\includegraphics{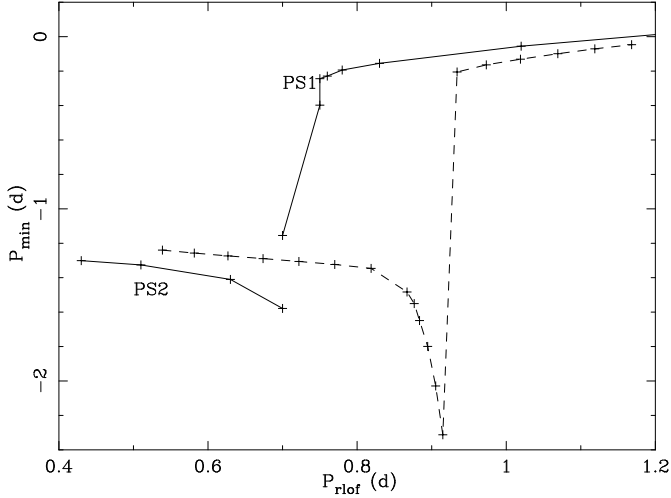}}}
 \caption{ 
 Comparison of our models (dashed line) to the models A25-I25 of PS1 and A25-Z25 
 of PS2 (solid lines).  See the text for details.
 }
 \label{fig:comp_ps}
\end{figure}

The question arises why  \citet{1988A&A...191...57P} and \citet{1989A&A...208...52P}, hereafter
PS1 and PS2, did not find ultra-compact systems in their study.  We tried to reproduce their
models with a 1.0\,$M_\odot$ compact primary and a 1.5\,$M_\odot$ secondary (models A25-I25 in PS1
and A25-Z25 in PS2) because these are best documented and they find the lowest minimum period here (38\,min for A25
in PS2). We calculated models with the same initial masses, mixing length ($l/H_\mathrm{p} = 1.5$), 
metallicity ($Z = 0.02$) and without 
overshooting. Figure~\ref{fig:comp_ps} compares their results to our calculations as the minimum period
($P_\mathrm{min}$) as a function of the period where Roche-lobe overflow starts ($P_\mathrm{rlof}$).

We find the bifurcation period at much larger $P_\mathrm{rlof}$, which is due to the fact that our stars rotate 
(about 25\% of the difference, according to test calculations we have done) and increased opacities.  Both effects increase the radii of our model stars, 
so that they must be placed in 
a larger orbit to fill their Roche lobe at the same stage of evolution.  It seems that shifting the two solid lines of PS1 and PS2
horizontally can approximately compensate for this, but the lines must be shifted over different amounts.  Hence,
a gap arises between what at first sight appears to be a continuous $P_\mathrm{rlof}$-range from PS1 and PS2.
The fact that they find the minimum period at the limit of each range, and the fact
that these two points are both at $P_\mathrm{rlof} = 0.70$\,d, but give very different minimum periods (100\,min and 38\,min
for PS1 and PS2 respectively), are supporting the evidence for the existence of this gap.  The cause for the gap
seems to be clear; in PS2 an accretion induced collapse (AIC) occurs when the primary reaches 1.44\,$M_\odot$, whereas
in PS1 no such event happens.  The AIC decreases the mass of the compact object and increases the orbital period
so that the further evolution can no longer be compared to that of systems without an AIC.  

In our more complete series of models, shown in Fig.~\ref{fig:comp_ps}, the lowest minimum
period we find is 7.0\,min, and is reached after 12.4\,Gyr.

\subsection{Comparison to Podsiadlowski et al.}
We chose the parameters of our models as similar as possible to those of 
\citet{2002ApJ...565.1107P} (see Sect.~\ref{sec:code}), to see if we could reproduce their results for a
1.4\,$M_\odot$ neutron star and a 1.0\,$M_\odot$ secondary. Indeed, the results
of our calculations are qualitatively very similar to their findings in their
Fig.~16 and their statement that binaries with an orbital period of 5\,minutes
can be achieved without a spiral-in, although we need slightly larger initial
periods to get to the same minimum period. \citet{2002ApJ...565.1107P} display their results as
a function of time since Roche-lobe overflow started, and because of this we cannot 
ascertain the total age of the binary at the minimum period.  
The red and blue model in their Fig.~16 reach minimum
periods of about 9 and 7 minutes, at approximately 4.5 and 5.5 Gyr after the
beginning of RLOF.  We find very similar results, and in addition we find the total ages of
these systems: 14 and 17\,Gyr respectively.  We find that it takes 13.4\,Gyr to
reach an orbital period of 11.4 minutes, the shortest period observed for an 
X-ray binary, and more than 35\,Gyr to shrink the orbit to 5\,minutes.  We conclude that it is
not possible to create systems with orbital periods less than 10\,min this way, within a Hubble time.

\citet{2002ApJ...565.1107P} find that there is a rather large range of initial
orbital periods (13 -- 17.7\,hr) that lead to a minimum period that is less than 
30\,minutes.  We find for the same condition a $P_\mathrm{rlof}$ of 15.3 -- 17.2\,hr,
which is considerably smaller.  This is firstly because our model stars have a slightly
larger radius.  Part of the explanation of the increased radius is given by the rotation of the star, 
although this can only account for 20\% of the difference in the $P_\mathrm{rlof}$-range, and by the different helium abundance 
(\citet{2002ApJ...565.1107P} use $Y = 0.27$, we have $Y = 0.26$), which explains 10\%.  
The larger radius shifts the whole $P_\mathrm{rlof}$-range to larger
orbital period.  Secondly, we limit our range to systems that reach their minimum period before 
the Hubble time, so that it is cut off above a certain $P_\mathrm{rlof}$.

What \citet{2002ApJ...565.1107P} call the {\it initial} period is the period at which RLOF initiates,
and which we call $P_\mathrm{rlof}$.  In the time before RLOF began, the magnetic braking may have played 
a role in shrinking the pre-RLOF orbit of the systems as listed in Table~\ref{tab:periods}.

\subsection{Comparison with observations and other models}

The main result from our computations is that, in a population where
all X-ray binaries evolve from close detached binaries of a
main-sequence star and a neutron star, systems with orbital periods
less than 30-40 minutes and with decreasing orbital periods are very
rare.  If we accept that the orbital period of the X-ray source in
NGC\,6624 is decreasing intrinsically (and not just observationally
due to gravitational acceleration), we must accept that it is a statistical fluke, 
or look for a different origin.

In this respect it would be important to know more about the orbital
periods and their derivatives of other X-ray sources in globular
clusters. 
A very short orbital period is detected for just one other bright
X-ray source, in NGC\,6712, as a regular variation of 0.044(7)\,mag in
one series of 53 F300W (wide U) filter HST observations with WFPC2 in
1995; aliasing allows two solutions at 13.2 or 20.6 minutes \citep{1996MNRAS.282L..37H}.
\citet{1996MNRAS.282L..37H} opt for the longer period, on the basis of the low
X-ray luminosity that reflects a low mass-transfer rate and a model
in which the donor to the neutron star is a white dwarf \citep{1987ApJ...312L..23V}.
We note that the same choice for the longer period would follow
for the magnetic-capture model.
The period derivative of the source in NGC\,6712 is not known. 
The argument that as many as half of the bright X-ray sources in
globular clusters have ultra-short periods is based on the similarity
of various properties of those X-ray sources with the properties of the
X-ray sources in NGC\,6624 and NGC\,6712. This argument is correct only if the
X-ray source in NGC\,6712 indeed has an ultra-short period. 
It is therefore important that this period is confirmed; which will
also settle between the aliases of 13.2 and 20.6 minutes.

Measurement of the period derivative will be very difficult. 
It is therefore of interest to know how many ultra-compact binaries
one would expect irrespective of their period derivative, in the
magnetic capture model. Alas, our computations stop a short time
after the minimum period, so that we do not have an accurate estimate
of the time spent at positive period derivative.
Nonetheless, inspection of our results as reflected in Fig.~\ref{fig:t-p_1.0}
shows that the evolution away from the minimum period
is only slightly slower than the evolution towards it.
Thus, the number of systems expected at the shortest
period range of between 10-30 minutes would only be a factor few
higher than the number in the same period range with decreasing period
only. This implies that the presence of even two systems with periods
less than 30 minutes among 13 globular cluster systems excludes the
magnetic-capture scenario as the dominant formation process. The
conclusion is true a fortiori if more such systems are discovered.

A donor in an ultra-compact system can also be a helium-burning
star. To bring such a small star into contact, a spiral-in must have
occurred \citep{1986A&A...155...51S}. 
The progenitor of such a helium-burning star would be more
massive than the main-sequence star found in globular clusters,
and \citet{1987ApJ...312L..23V} argued that this excludes such donors for
sources in globular clusters. However, more massive stars can be made
in direct collisions: if such a more massive star ends up in a binary with a 
neutron star, further evolution can lead to a helium-burning
donor in an ultra-compact system. This scenario may gain in importance
if tidal capture is indeed less efficient, as indicated by a high
fraction of systems with ultra-short periods. It allows negative
derivatives of the orbital period.

Since the measurement of the intrinsic derivative of the orbital period
is so difficult, it is useful to look for other observational properties
that can discriminate between the different origins of an ultra-compact
binary. With this in mind, we refer to Table~\ref{tab:properties} where some properties
of ultra-short-period systems are listed that follow for the magnetic-capture model,
in particular the mass-transfer rate at various periods, and the
abundances of the more important elements.
A pure white-dwarf donor, whittled down to a mass less than 0.1\,$M_\odot$,
would have no hydrogen if it was a helium white dwarf; and no hydrogen and
no helium if it was a carbon-oxygen white dwarf.
Therefore, if hydrogen is discovered in the spectrum of an ultra-compact X-ray binary,
this indicates evolution through magnetic capture and the orbital period must still
be decreasing.  Close to the minimum period the hydrogen abundance at the surface goes 
to zero and thus is no longer discriminant between models.

\begin{acknowledgements}
We thank P.P. Eggleton for making available to us the latest version of his binary evolution code.
We also thank F. Rasio for his suggestions to improve this article.
\end{acknowledgements}

\bibliographystyle{aa}
\bibliography{1777}

\end{document}